\documentclass[sigconf]{acmart}

\usepackage{booktabs}
\usepackage{multirow}
\usepackage{threeparttable}

\usepackage{graphicx}
\usepackage{subcaption}
\usepackage{algorithmic}
\usepackage{algorithm}
\usepackage{multirow}
\usepackage{amsmath}

\usepackage[normalem]{ulem}
\useunder{\uline}{\ul}{}

\AtBeginDocument{%
  }

\acmVolume{0}
\acmNumber{0}
\acmArticle{0}
\acmMonth{0}

\begin{document}

\title{Generalizable and Interpretable RF Fingerprinting with Shapelet-Enhanced Large Language Models}


\author{Tianya Zhao}
\email{tzhao010@fiu.edu}
\orcid{0000-0002-3808-7549}
\affiliation{%
  \institution{Florida International University}
  \city{Miami}
  \state{Florida}
  \country{USA}
}

\author{Junqing Zhang}
\email{junqing.zhang@liverpool.ac.uk}
\affiliation{%
  \institution{University of Liverpool}
  \city{Liverpool}
  \country{UK}
}

\author{Haowen Xu}
\email{hxu4@wpi.edu}
\affiliation{%
  \institution{Worcester Polytechnic Institute}
  \city{Worcester}
  \state{Massachusetts}
  \country{USA}
}

\author{Xiaoyan Sun}
\email{xsun7@wpi.edu}
\affiliation{%
  \institution{Worcester Polytechnic Institute}
  \city{Worcester}
  \state{Massachusetts}
  \country{USA}
}

\author{Jun Dai}
\email{jdai@wpi.edu}
\affiliation{%
  \institution{Worcester Polytechnic Institute}
  \city{Worcester}
  \state{Massachusetts}
  \country{USA}
}

\author{Xuyu Wang}
\email{xuywang@fiu.edu}
\affiliation{%
  \institution{Florida International University}
  \city{Miami}
  \state{Florida}
  \country{USA}
}

\begin{abstract}
Deep neural networks (DNNs) have achieved remarkable success in radio frequency (RF) fingerprinting for wireless device authentication. However, their practical deployment faces two major limitations: domain shift, where models trained in one environment struggle to generalize to others, and the black-box nature of DNNs, which limits interpretability. To address these issues, we propose a novel framework that integrates a group of variable-length two-dimensional (2D) shapelets with a pre-trained large language model (LLM) to achieve efficient, interpretable, and generalizable RF fingerprinting. The 2D shapelets explicitly capture diverse local temporal patterns across the in-phase and quadrature (I/Q) components, providing compact and interpretable representations. Complementarily, the pre-trained LLM captures more long-range dependencies and global contextual information, enabling strong generalization with minimal training overhead. Moreover, our framework also supports prototype generation for few-shot inference, enhancing cross-domain performance without additional retraining. To evaluate the effectiveness of our proposed method, we conduct extensive experiments on six datasets across various protocols and domains. The results show that our method achieves superior standard and few-shot performance across both source and unseen domains. 
\end{abstract}



\received{20 February 2007}
\received[revised]{12 March 2009}
\received[accepted]{5 June 2009}

\maketitle

\section{Introduction}\label{sec:intro}

The rapid growth of the Internet of Things (IoT) has led to the ubiquitous integration of wireless technologies in daily life. As a result, robust device authentication is essential to ensure secure access for legitimate users while blocking malicious ones. Traditional cryptographic methods, such as those relying on Internet Protocol (IP) or Media Access Control (MAC) addresses~\cite{lehtonen2009securing}, are commonly used but remain vulnerable to spoofing and tampering~\cite{zou2016survey}. Additionally, these methods may not suit ultra-low-power devices or outdated, unmaintained hardware~\cite{formby2016s}. To overcome these limitations, radio frequency (RF) fingerprinting offers a compelling solution that exploits device-specific characteristics to enable reliable identification and enhanced security across various applications.

RF fingerprints result from minute physical imperfections in the analog circuitry of the device during the manufacturing process~\cite{zhang2025physical}. These subtle imperfections slightly affect transmitted signals without compromising overall device functionality, resulting in a distinct fingerprint for each RF emitter, including ultra-low-power and legacy devices. The existing RF fingerprinting can be categorized into traditional extractor-based and deep learning-based. Traditional methods require manually designed feature extractors to capture hardware characteristics, which can be complex, demand extensive protocol knowledge, and heavily depend on the extraction algorithm. In contrast, deep learning-based methods automate feature extraction and classification, leveraging raw in-phase and quadrature (I/Q) samples to simplify the process and enhance accuracy~\cite{zhang2025physical}.

While deep neural networks (DNNs) have been extensively employed to extract and classify RF fingerprints with high accuracy, current approaches face two critical limitations: \textit{domain shift} and \textit{lack of interpretability}. Domain shift occurs when a model trained in one domain (e.g., specific location, environment, or time period) performs poorly in a different one, limiting its generalization across diverse real-world scenarios. To mitigate this, techniques like adversarial domain adaptation~\cite{li2022radionet}, few-shot learning (FSL)~\cite{zhao2024few}, and self-supervised learning (SSL)~\cite{liu2021self} have been widely explored to enhance model generalization and adaptability. However, domain adaptation typically relies on massive labeled data or domain-specific information, which are often difficult and time-consuming to obtain in RF fingerprinting due to the labor-intensive process of signal annotation. 
Furthermore, both FSL and domain adaptation require complex training strategies, such as adversarial alignment or episode-based training~\cite{snell2017prototypical}, which complicate deployment and may limit scalability.

In contrast, SSL simply leverages unlabeled data to pre-train models, learning features that are robust to domain shifts while avoiding costly annotations and complex training strategies. This makes SSL particularly suitable for RF fingerprinting~\cite{liu2023overcoming}. Building on the strengths of SSL, recent advances in Large Language Models (LLMs), such as BERT and the GPT series, have demonstrated remarkable generalization capabilities across diverse tasks and domains. While the integration of LLMs into RF fingerprinting remains relatively underexplored, their potential in wireless applications is evident. For example, WirelessLLM~\cite{shao2024wirelessllm} empowers LLMs with knowledge and expertise in wireless communication, while RFSensingGPT~\cite{khan2025rfsensinggpt} develops an integrated LLM-based retrieval system for RF technical contents. These LLMs offer significant benefits in areas such as code generation, domain-specific reasoning, and spectrum analysis. Despite these strengths, their performance in classification tasks remains limited, possibly due to the difficulty of aligning RF data with language prompts and the absence of RF-specific pre-training.

The second limitation in current DNN-based approaches is their lack of interpretability. This is especially critical in safety-sensitive applications such as RF fingerprinting, where understanding model behavior can enhance system reliability. Although post-hoc interpretability techniques exist, they typically require additional processing and cannot provide end-to-end interpretability. While recent prompting strategies like Chain-of-Thought (CoT) can reveal intermediate reasoning steps in LLMs, these methods are primarily tailored to symbolic or linguistic reasoning and have limited applicability to RF signals. Although \cite{zhao2024cross} integrates an explanation module for RF fingerprinting, its purpose is primarily to augment data and boost performance, rather than to provide explicit insights into the model’s internal reasoning. Overall, these limitations highlight the need to explore a new method that can integrate LLMs into RF fingerprinting pipelines while balancing generalization, efficiency, and interpretability.

\textbf{Challenges.} 
Integrating powerful LLMs into RF fingerprinting systems to address domain shift while maintaining a balance between performance and interpretability presents several challenges.
First, LLMs are primarily trained on textual data and lack an inherent understanding of the unique characteristics of RF signals. While LLMs have shown promise in time series analysis~\cite{jiang2024empowering}, their effectiveness in RF fingerprinting is not guaranteed due to significant domain shifts and the complex, domain-specific nature of RF data, which can impair performance and generalization. Therefore, the primary challenge is how to leverage the generalization power of pre-trained LLMs for cross-domain RF fingerprinting without extensive retraining.
Second, LLMs are proficient in few-shot inference, where they can complete tasks using only one or a few examples provided in the prompt. However, applying this strength to RF fingerprinting is non-trivial, given the high variability of RF data and its structural dissimilarity to text-based inputs.
Third, for safety-critical applications like RF fingerprinting, intrinsic interpretability is highly desirable. Built-in explanation mechanisms within DNNs are often more reliable and efficient than post-hoc interpretability techniques. However, embedding intrinsic interpretability into large and complex LLMs without cumbersome training processes or compromising performance remains a significant challenge.

\textbf{Solution.}
To address these challenges, we carefully adapt pre-trained LLMs to improve generalization and reduce training costs, and propose a learnable shapelets module to provide interpretability. 
Rather than using raw RF data directly as input prompts, we employ the LLM as a feature extractor to obtain robust and discriminative global features from I/Q data. This design is supported by~\cite{zhou2023one}, showing that LLMs theoretically and empirically perform functions similar to principal component analysis (PCA) and outperform various DNNs on different time series tasks. To bridge the modality gap, an input embedding module is employed to project I/Q data to the input space of LLMs. To preserve the pre-trained knowledge and reduce training cost, we freeze most parameters, updating only the positional embeddings and layer normalization during training.
To leverage the few-shot inference capabilities of LLMs, we adapt the concept of prototypical network (PTN)~\cite{snell2017prototypical} to create class-specific prototypes that enable efficient few-shot inference without requiring retraining.
For interpretability, we integrate variable-length two-dimensional (2D) shapelets that explicitly capture fine-grained local patterns within RF data. These shapelets are integrated into the model to highlight discriminative subsequences, offering built-in explanations for classification decisions and enhancing both performance and transparency.
The main contributions of this paper are as follows.
\begin{itemize}
    \item To the best of our knowledge, this is the first work to explore the integration of pre-trained LLMs into RF fingerprinting systems to enhance generalization in cross-domain and cross-dataset scenarios.
    \item We propose a novel interpretable fine-tuning framework that adapts a pre-trained LLM with variable-length 2D learnable shapelets for the RF fingerprinting task. This approach offers built-in interpretability without the need for computationally intensive retraining, while preserving the generalization capabilities of the pre-trained LLM.
    \item We conduct comprehensive experiments on various protocols, including Wi-Fi, LoRa, and Bluetooth Low Energy (BLE), across multiple datasets and scenarios. The superior results show the broad applicability and effectiveness of our method in addressing domain shift.
\end{itemize}

The rest of the paper is organized as follows. Section~\ref{sec:relatedwork} discusses the related work, and Section~\ref{sec:pre} introduces the preliminary. Section~\ref{sec:formulation} illustrates the problem formulation of this study. Our methodology is introduced in Section~\ref{sec:method}. In Section~\ref{sec:ex_eva}, we conduct comprehensive experimental evaluations. Section~\ref{sec:future} gives limitations and future work. Section~\ref{sec:conclusion} concludes this paper.

\section{Related Work}\label{sec:relatedwork}
Domain shift presents a major challenge for wireless systems, as variations in the environment and temporal drift can lead to substantial drops in accuracy when models are applied to previously unseen domains~\cite{al2020exposing, jagannath2023embedding, yuan2025robust}. To address this issue, several strategies are commonly employed, including data augmentation, domain adaptation, FSL, and SSL~\cite{zhou2022domain}. 
Data augmentation techniques are widely used to improve model generalization by enriching the diversity of the training dataset. For instance, DeepLoRa employs channel model-based data augmentation to improve the robustness of LoRa fingerprinting~\cite{al2021deeplora}. Wang \textit{et al.}~\cite{wang2024ai} propose a modified generative model to synthesize I/Q samples, thereby improving classification accuracy in satellite fingerprinting tasks. In terms of domain adaptation, RadioNet~\cite{li2022radionet} employs adversarial learning and a novel metric to improve performance under cross-day scenarios. Pan \textit{et al.}~\cite{pan2024equalization} integrate channel equalization to further boost adaptation capability in RF fingerprinting tasks.

FSL and SSL have also emerged as promising solutions to address domain shift, especially under limited supervision or labeled data. Yao \textit{et al.}~\cite{yao2023few} adopt an asymmetric masked auto-encoder within an FSL framework for specific emitter identification (SEI). Zhao \textit{et al.}~\cite{zhao2024few} combine PTN with data augmentation to improve generalization in unmanned aerial vehicle authentication. Similarly, Zhang \textit{et al.}~\cite{zhang2022data} propose different data augmentations to support FSL for SEI. SSL is employed in~\cite{liu2023overcoming} as a complementary strategy to FSL, effectively reducing the need for labeled data in the unseen target domain. Chen \textit{et al.}~\cite{chen2024unsupervised} adopt contrastive learning to extract domain-invariant features, demonstrating its effectiveness in mitigating domain-specific variations for robust RF fingerprinting. Li \textit{et al.}~\cite{li2024self} propose a momentum-based asymmetric SSL method to enhance feature extraction capability for SEI.

The most closely related study is~\cite{zhao2024cross}, which incorporates eXplainable AI (XAI) for data augmentation to fine-tune the feature extractor, thereby enhancing target domain performance within the FSL framework. However, their approach does not exploit the advantages of SSL, and the use of interpretation is implicit, serving only as a tool for data augmentation rather than understanding model behavior.
In contrast, our work differs from related studies in several key aspects. First, we provide end-to-end and intrinsic explanations that can bring meaningful insights into model behavior beyond data augmentation.
Second, we integrate LLMs to leverage their powerful generalization capabilities to enhance the robustness of the learned representations. Third, we exploit the few-shot inference to enable effective generalization across previously unseen domains without retraining.


\section{Preliminary}\label{sec:pre}

\subsection{RF Fingerprinting}
Fingerprinting has been extensively studied as a physical-layer authentication technique for securing IoT devices in wireless networks. Compared with cryptographic or protocol-level identifiers, physical-layer authentication relies on intrinsic hardware or behavioral characteristics that are difficult to forge and can be captured passively without protocol modifications or additional energy overhead~\cite{he2025ht, han2018butterfly, lee2019voltkey, wang2024earslide, wang2022toothsonic}. These properties make it particularly well-suited for low-power, legacy, and resource-constrained systems.

In fingerprinting, device-specific imperfections introduced during hardware manufacturing manifest as subtle but distinctive patterns in physical signals, enabling reliable device identification. A variety of modalities have been explored, including RF signals~\cite{zhao2024cross, li2022reliable}, electromagnetic emissions~\cite{feng2023fingerprinting, lee2022aerokey, shen2022electromagnetic}, and magnetic side channels~\cite{cheng2019demicpu}.

In this paper, we focus on RF fingerprinting for various IoT protocols, where device identity is inferred from raw or lightly processed RF signals collected at the receiver. Due to unavoidable hardware imperfections, such as oscillator instability, power amplifier nonlinearity, and I/Q imbalance, signals transmitted by different devices exhibit unique characteristics, even when transmitting identical payloads~\cite{zhang2025physical}. These device-dependent patterns form the basis of RF fingerprints.

\subsection{Large Language Model}
LLMs are a class of DNNs trained on large-scale corpora using SSL objectives, enabling them to learn powerful representations of sequential data without the need for manual annotation. Models such as BERT~\cite{devlin2018bert} and GPT~\cite{radford2019language}, built on the Transformer architecture, leverage self-attention mechanisms to model contextual dependencies over long sequences. Each Transformer layer includes multi-head self-attention, a feed-forward network, residual connections, and layer normalization for stable training. The architecture of existing LLMs can be categorized into encoder-only (e.g., BERT), decoder-only (e.g., GPT), or encoder-decoder (e.g., T5) configurations. 
Beyond their success in natural language processing, recent research has shown that LLMs exhibit strong capabilities in transfer learning, few-shot adaptation, and representation learning over other modalities~\cite{xu2021limu}. This motivates the exploration of LLMs in non-linguistic tasks such as RF fingerprinting, potentially benefiting from the LLMs’ ability to extract robust and discriminative patterns for enhanced generalization across different domains.

\subsection{Shapelet}\label{subsec:shapelet}

Shapelets are discriminative subsequences derived from time-series data that effectively capture characteristic patterns essential for classification tasks~\cite{ye2009time}. Unlike traditional whole-series similarity methods, such as Dynamic Time Warping (DTW), which are computationally intensive and often less accurate~\cite{bagnall2017great}, shapelet-based approaches identify local, class-specific patterns. These subsequences serve as interpretable features, enabling high classification accuracy while providing human-understandable insights into the model's decision-making process. In the context of RF fingerprinting, shapelets can identify unique temporal patterns in RF signals that distinguish one device from another.

Given a univariate time-series dataset, the input space is defined as $\mathcal{X} \subset \mathbb{R}^{\mathit{T}}$, where $\mathit{T} \in \mathbb{N}$ is the length of each instance. For the $i$-th instance $\mathbf{x}_{i}$, a subsequence $\mathbf{x}_{i,j} \in \mathbb{R}^L$ starting at time index $j$ is defined as:
\begin{equation}
    \mathbf{x}_{i,j} = (x_{i,j}, \ \dots, \ x_{i,j+L-1}), \quad 1 \leq j \leq \mathit{J},
\end{equation}
where $\mathit{L}$ is the length of the subsequence and $\mathit{J} = \mathit{T} - \mathit{L} + 1$ is the total number of subsequences that can be extracted from a single instance.
A shapelet is a subsequence of length $\mathit{L}_s$ with strong discriminative power for classification. While shapelets are essentially subsequences by definition, only those with significant class-separating properties are chosen as shapelets. To discover such shapelets, existing methods can be broadly categorized into search-based and learning-based approaches. Search-based methods typically conduct exhaustive or randomized searches of all possible subsequences in the training data~\cite{hills2014classification, lines2012shapelet}. In contrast, learning-based methods treat shapelets as continuous, learnable parameters and optimize them jointly with a classifier~\cite{grabocka2014learning, yamaguchi2023time}.

\section{Problem Formulation}\label{sec:formulation}

DNN-based RF fingerprinting leverages DNNs to automatically extract unique signal features as fingerprints from raw I/Q samples for device identification. 
Therefore, RF fingerprinting systems should remain both reliable and robust for practical deployment~\cite{xu2015device}. This paper focuses on two fundamental objectives in DNN-based RF fingerprinting: \textit{generalization} and \textit{interpretability}. Specifically, our goals are (1) enhancing model generalization across diverse domains, and (2) providing intrinsic interpretability of model decisions without compromising classification performance.

Let the source RF fingerprinting dataset be denoted as $\mathcal{D} = \{(\mathbf{x}_i, y_i)\}_{i=1}^N$, where the input space is defined as $\mathcal{X} \subset \mathbb{R}^{2 \times T}$ and the label space as $\mathcal{Y}=\{1, \dots, C\}$.
Each input $\mathbf{x}_i \in \mathcal{X}$ represents a real-valued matrix constructed by separating the in-phase (I) and quadrature (Q) components of a complex RF signal into two channels of length $T$. The label $y_i \in \mathcal{Y}$ denotes the device identity among $C$ unique transmitters. 
Given that labeled data from new domains (e.g., deployment locations or communication conditions) is hard to obtain, the \textit{generalization objective} is to train a model $f: \mathcal{X} \rightarrow \mathcal{Y}$ that not only perform well on the source domain but also generalize effectively to unseen target domain data $\mathcal{D}' = \{(\mathbf{x}_i', y_i')\}_{i=1}^{N'}$, using little or no labeled data.
This can be formalized as a constrained optimization problem:
\begin{equation}
    \min_{f} \ \mathcal{E}_{\text{target}} \quad \text{s.t.} \quad \mathcal{E}_{\text{source}} \leq \epsilon,
\end{equation}
where $\mathcal{E}_{\text{source}}$ and $\mathcal{E}_{\text{target}}$ denote the expected error on the source and target domain data, respectively. The constraint $\mathcal{E}_{\text{source}} \leq \epsilon$ ensures strong performance on the source domain.

The \textit{interpretability objective} is to construct an interpretable model $g$ that approximates the original model $f$ while providing intrinsic explanations for its predictions. To ensure that interpretability does not compromise utility, we require that the performance gap between the interpretable model $g$ and the original model $f$ remains within a small acceptable threshold: $|\mathcal{E}_{g} - \mathcal{E}_{f}| \leq \delta$, where $\mathcal{E}_g$ and $\mathcal{E}_f$ denote the expected error of the interpretable and original models, respectively. The constant $\delta$ denotes a small, acceptable level of performance degradation introduced by enhancing interpretability.

\section{Methodology}\label{sec:method}

\subsection{Overview}

\begin{figure}[ht]
\makebox[\columnwidth][c]
{\includegraphics[width=\columnwidth]{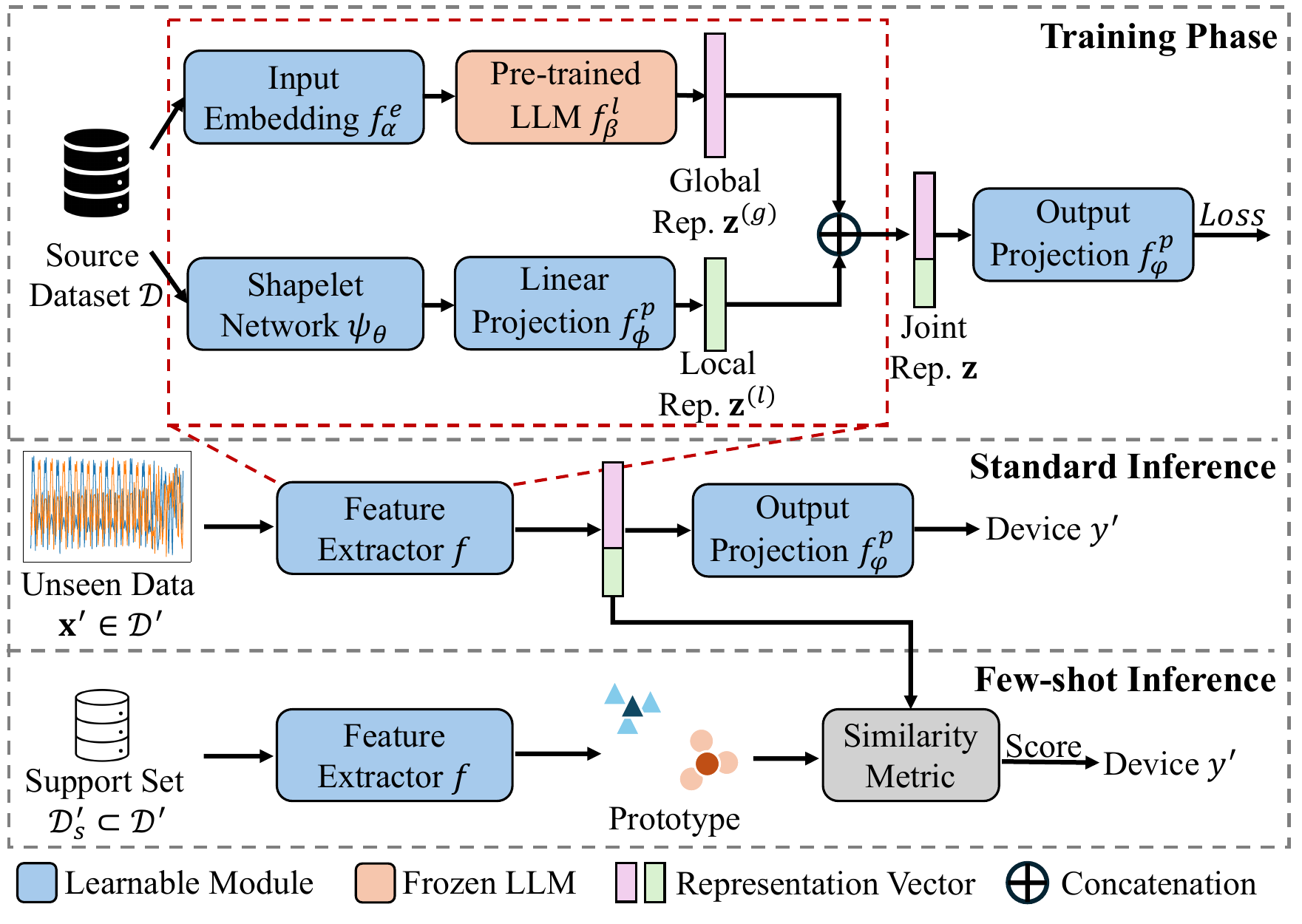}}
    \caption{Overview of the proposed RF fingerprinting system. Few-shot inference is enabled when target domain data is available.}
    \label{fig:rff_system}
\end{figure}

In this paper, we aim to leverage the generalization capability of pre-trained LLMs to improve cross-domain performance and build an intrinsic interpretable shapelet model to provide explanations for the RF fingerprinting system. The overview of our proposed method is shown in Fig.~\ref{fig:rff_system}. 
First, we employ an input embedding module to project I/Q data into the dimensional space of the specific LLM, allowing it to model long-range dependencies and global contextual information. In parallel, a shapelet network explicitly captures local signal characteristics for interpretation. 
These global and local representations are then concatenated to form a joint representation, which is further refined through an output projection module for classification.
During inference, unseen I/Q data is processed to extract a joint representation.
For standard inference, the joint representation is directly projected to the device label space. For the few-shot inference scenario, the system compares the joint representation against class prototypes derived from a small support set $\mathcal{D}'_s \subset \mathcal{D}'$, using similarity-based matching to identify the target device.

\subsection{Pre-trained LLM Adaptation}
To leverage the strong generalization ability of pre-trained LLMs for addressing domain shift in RF fingerprinting, the critical step is to adapt these language-centric models to RF data and enable them to effectively capture robust representations. To this end, we propose an input embedding module tailored for RF data and a lightweight fine-tuning strategy to align LLMs with the statistical characteristics of RF data.

\subsubsection{RF Input Embedding}
Pre-trained LLMs typically expect input as a sequence of token embeddings, where each token is represented by a fixed-dimensional vector of size $d_h$, commonly referred to as the hidden size. For instance, models such as GPT-2 require each token to lie in a $768$-dimensional space, resulting in an input of shape $l_{seq} \times d_{h}$, where $l_{seq}$ denotes the input sequence length. In contrast, raw I/Q data is a 2D time-series signal that is structurally incompatible with this format. To bridge this gap, we introduce an input embedding module $f_{\alpha}^{e}:\mathbb{R}^{2\times T} \rightarrow \mathbb{R}^{l_{seq} \times d_{h}}$, which transforms I/Q data into fixed-dimensional embeddings compatible with the LLM's input requirements.

A common approach in time-series modeling is to segment the input into patches~\cite{nie2022time}, and use a linear projection to make the data fit the expected input structure. However, this manual segmentation can potentially split a single RF fingerprint feature across multiple patches, thereby degrading representational integrity. To mitigate this, we employ a lightweight convolutional neural network (CNN)-based encoder as the input embedding module $f_{\phi}^{e}$. This design choice exploits the CNN's proven effectiveness in capturing local dependencies and spatially coherent features in RF fingerprinting tasks~\cite{zhao2024cross, sankhe2019oracle, al2020exposing}. The CNN encoder processes raw I/Q data into a structured sequence of embeddings $f_{\alpha}^{e}(\mathbf{x})$, which is then fed directly into the LLM as its tokenized input. This embedding strategy preserves fine-grained temporal structures essential for device-level identification while enabling the model to leverage the generalization capabilities of pre-trained LLMs.

\begin{figure}[ht]
    \centering
    \subfloat[Raw input]{%
        \includegraphics[width=0.32\columnwidth]{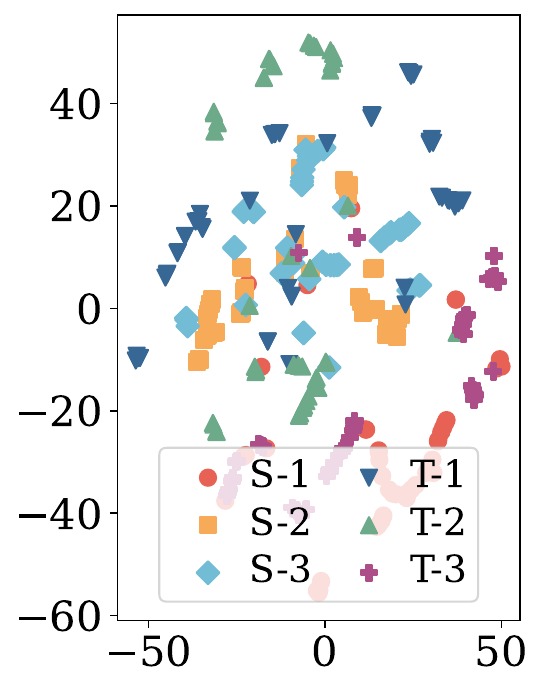}%
        \label{fig:tsne1}
    }
    \subfloat[Frozen LLM]{%
        \includegraphics[width=0.32\columnwidth]{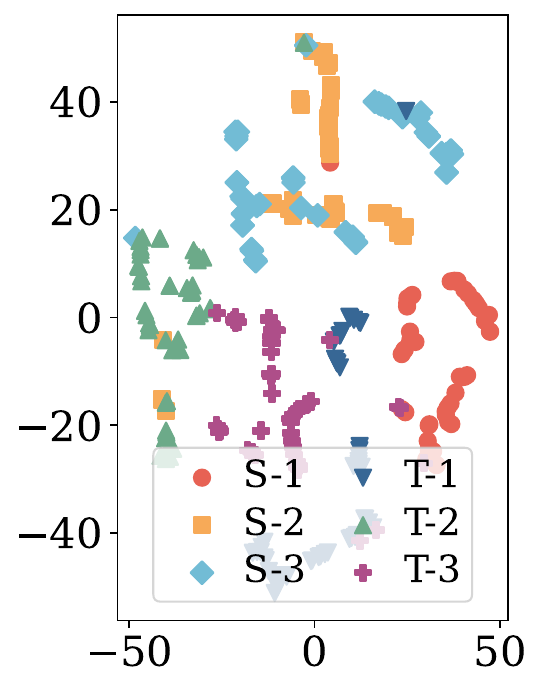}%
        \label{fig:tsne2}
    }
    \subfloat[Unfrozen layer norm]{%
        \includegraphics[width=0.32\columnwidth]{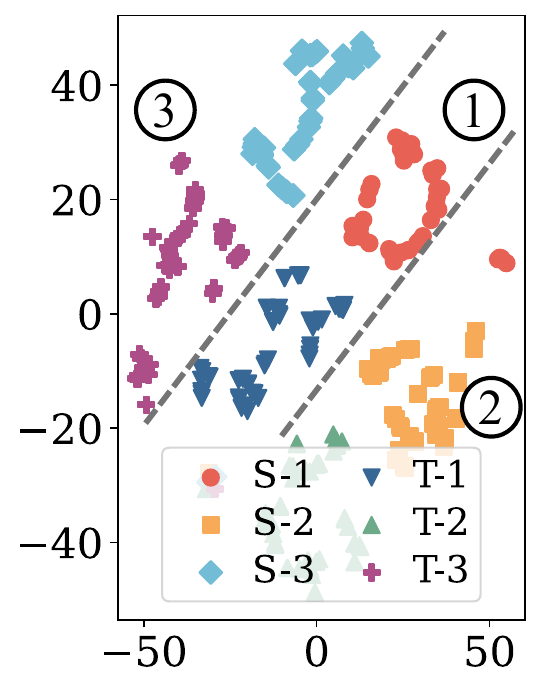}%
        \label{fig:tsne3}
    }
    \caption{The t-SNE visualization of features under different LLM adaptation strategies. S-$1$ to S-$3$ represent three devices in the source domain, while T-$1$ to T-$3$ denote the same devices in the target domain.}
    \label{fig:tsne_all}
\end{figure}


\subsubsection{Frozen LLM Fine-tuning}
Training LLMs from scratch for RF fingerprinting is computationally prohibitive due to their massive parameter sizes and extensive training requirements. Instead, we leverage the generalization capabilities of pre-trained LLMs to extract robust representations from I/Q data while avoiding the cost of full-model retraining. Since the majority of the learned knowledge in LLMs is encoded within the self-attention layers and feedforward networks~\cite{zhou2023one}, we choose to freeze these components during fine-tuning to preserve their generalization capabilities. To further align the model with the distributional characteristics of RF data, we only fine-tune the parameters of the positional embedding and layer normalization components. This allows for minimal yet effective adaptation to the statistical properties of RF signals without disrupting the LLM's pre-trained knowledge.

As illustrated in Fig.~\ref{fig:tsne_all}, we visualize t-SNE embeddings for three devices each from the source (S-$1$ to S-$3$) and target (T-$1$ to T-$3$) domains. In Fig.~\ref{fig:tsne1}, raw samples from different devices and domains are heavily entangled, reflecting a significant domain shift. In Fig.~\ref{fig:tsne2}, features extracted by a fully frozen BERT model form more structured clusters, though the separation between domains remains limited. In contrast, Fig.~\ref{fig:tsne3} shows that fine-tuning only the layer norm components results in well-separated clusters, both across domains and devices. Specifically, the clusters at the top correspond to device $3$, those in the middle to device $1$, and those at the bottom to device $2$, highlighting the model’s improved ability to distinguish devices across domains.

Formally, let $f_{\beta}^{l}$ denote a pre-trained LLM, where only the layer normalization parameters $\beta$ are fine-tuned during adaptation. Given an input RF sample $\mathbf{x}_i$ in the I/Q domain, we first extract a local embedding using a CNN-based encoder $f_{\alpha}^{e}$. 
This embedding is then fed into the LLM to obtain a global feature vector $\mathbf{z}^{(g)}$ that summarizes long-range dependencies and global contextual information:
\begin{equation}
    \mathbf{z}_{i}^{(g)} = f_{\beta}^{l}(f_{\alpha}^{e}(\mathbf{x}_i)), \quad \mathbf{z}_{i}^{(g)} \in \mathbb{R}^{d_{h}}.
\end{equation}
By combining a CNN-based input embedding with a frozen LLM that only fine-tunes layer normalization, our method efficiently adapts the LLM to the RF domain and provides a solid foundation for learning robust representations.


\subsection{Learnable 2D Shapelets}
Interpretability in RF fingerprinting models is essential for understanding how specific signal features contribute to classification decisions. In time-series tasks, shapelet-based methods are widely used to extract discriminative subsequences that offer insight into model behavior~\cite{bagnall2017great}. However, traditional shapelets are typically one-dimensional (1D) and fixed in length, which may limit their effectiveness in modeling the complex and variable patterns in RF data. This is because both I and Q components carry the information of the signal, and they often exhibit discriminative patterns through their joint behavior, such as synchronized amplitude and phase shifts. 
In addition, the fixed-length constraint of traditional shapelets limits their ability to model patterns that emerge over diverse temporal scales. Critical device signatures may appear in short bursts or over longer intervals; fixed-length shapelets may either miss long-term dependencies or include excessive noise from irrelevant subsequences, thereby degrading both interpretability and model accuracy.

To address these limitations, we propose a novel interpretable module that learns a group of variable-length 2D shapelets to effectively capture explicit and discriminative local temporal patterns in I/Q data, while providing interpretable insights into the learned representations. Each shapelet jointly spans both signal components and is optimized end-to-end with the pre-trained LLM.
Formally, the 2D shapelet group is defined by a configuration set $\{(M_1, L_1), \dots, (M_m, L_m)\}$, where $M_i$ represents the number of shapelets with length $L_i$ for the $i$-th group, and $i=1, \cdots, m$. For clarity, we index all shapelets as $\{\mathbf{S}_{k}\}_{k=1}^K$, where $K = \sum_{i=1}^{m}M_i$, and each $\mathbf{S}_{k} \in \mathbb{R}^{2 \times L_k}$ represents a 2D subsequence that captures local patterns across the I and Q components over a window of length $L_k$.

These 2D shapelets are implemented as learnable parameters within a shallow neural module $\psi_{\theta}$, referred to as the Shapelet Network in Fig.~\ref{fig:rff_system}. This network is trained end-to-end with the rest of the model through backpropagation, allowing the shapelets to automatically adapt to the most discriminative patterns in the data.
The shapelet matching process begins by extracting subsequences from the input I/Q data using a sliding window. For each shapelet $\mathbf{S}_{k} \in \mathbb{R}^{2 \times L_k}$, we extract all possible subsequences of length $L_k$ from the input $\mathbf{x}_i \in \mathbb{R}^{2 \times T}$. Unlike the univariate subsequence defined in Section~\ref{subsec:shapelet}, the $j$-th subsequence for the $i$-th I/Q data $\mathbf{x}_i$ is defined as:
\begin{equation}
    \mathbf{x}_{i,j}^k = (\mathbf{x}_{i,t_j}, \ \dots, \ \mathbf{x}_{i,t_j+L_k-1}), \quad 1 \leq j \leq J_k,
\end{equation}
where $J_k = T-L_k+1$ represents the total number of I/Q subsequences, and $\mathbf{x}_{i,j}^k$ denotes the specific subsequence being compared against the shapelet $\mathbf{S}_{k}$.

We measure the similarity between each shapelet and the input data by computing distances to all extracted subsequences. Following previous work~\cite{ye2009time}, the distance between the input data $\mathbf{x}_i$ and the $k$-th shapelet $\mathbf{S}_{k}$ is defined as the minimum distance between the shapelet and all input data subsequences. Intuitively, this distance measures how well the shapelet matches the most similar local pattern in the signal. In this work, we use the Euclidean distance metric, defined as $d_{i,j}^k =  \left\| \mathbf{S}_{k} - \mathbf{x}_{i,j}^k \right\|_2,$ where $d_{i,j}^k$ quantifies the distance between the shapelet $\mathbf{S}_k$ and the $j$-th subsequence $\mathbf{x}_{i,j}^k$. 
To enable differentiable learning, we compute a soft activation score by applying softmax pooling across all subsequences. Specifically, the activation of shapelet $\mathbf{S}_k$ for $\mathbf{x}_i$ is defined as:
\begin{equation}
a_{i}^k = \sum_{j=1}^{J_k} w_{i,j}^k \cdot (-d_{i,j}^k), \quad
w_{i,j}^k = \frac{\exp(-d_{i,j}^k)}{\sum_{j'=1}^{J_k} \exp(-d_{i,j'}^k)},
\end{equation}
where $w_{i,j}^k$ denotes the weight assigned to the $j$-th subsequence. The negative distance ensures that smaller distances result in higher activation values.
Then, we obtain a full shapelet activation vector $\psi_{\theta}(\mathbf{x}_i):=\mathbf{a}_i=[a_{i}^1, \dots, a_{i}^K] \in \mathbb{R}^K$, where each element indicates the matching strength between the input and a specific shapelet. This vector is passed to a linear projection $f_{\phi}^p$ to produce the local representation:
\begin{equation}
    \mathbf{z}_i^{(l)} = f_{\phi}^p(\psi_{\theta}(\mathbf{x}_i)), \quad \mathbf{z}_i^{(l)} \in \mathbb{R}^{d_l}.
\end{equation}

\subsection{Joint Representation}
To effectively capture both global context and local discriminative features for RF fingerprinting, we combine the global feature vector $\mathbf{z}^{(g)} \in \mathbb{R}^{d_h}$, derived from a frozen pre-trained LLM, with the local representation $\mathbf{z}^{(l)} \in \mathbb{R}^{d_l}$, produced by the shapelet network. These representations are concatenated to form a joint representation:
\begin{equation}
    \mathbf{z} = \mathbf{z}^{(g)} \oplus \mathbf{z}^{(l)}, \quad \mathbf{z} \in \mathbb{R}^{d_h+d_l},
\end{equation}
where $\oplus$ denotes concatenation. This joint representation integrates complementary information across semantic and temporal dimensions, enabling a more comprehensive characterization of device-specific RF signals. The joint representation is then passed through a learnable output projection module $f_\varphi^p$ to produce the final logits for RF fingerprinting.

\subsection{Loss Function}
To train the proposed framework, we adopt a composite loss function that promotes accurate classification and interpretable shapelet-based representations.
The model is first optimized by the standard cross-entropy loss $\mathcal{L}_{\text{cls}}=-\sum_{i=1}^{N} \log p(y_i|f_\varphi^p(\mathbf{z}_i))$, where $p(y_i|\cdot)$ is the softmax probability over device labels, updating the shapelet network and the unfrozen parts of the pre-trained LLM, encouraging effective class discrimination from the joint representation.
However, relying solely on cross-entropy loss does not constrain the behavior of shapelet activations in the shapelet network. Without additional regularization, shapelet responses may become dense and highly correlated, with multiple shapelets activating similarly across different input instances. This reduces interpretability by making it difficult to identify which shapelets capture unique temporal patterns and diminishes the effectiveness of the local representation by introducing redundancy.

To address this, we introduce two regularization objectives: sparsity and diversity, to promote meaningful and non-redundant shapelet activation patterns.
Intuitively, we expect each input instance to be represented by only a few highly relevant shapelets, rather than activating all shapelets uniformly. This encourages each shapelet to specialize in capturing distinct patterns. To enforce this behavior, we apply an $l_1$-based sparsity regularization $\mathcal{L}_{\text{spr}} = \frac{1}{N} \sum_{i=1}^{N} ||\mathbf{a}_i||_1,$
which penalizes large or dense activations and promotes interpretability by allowing a small subset of shapelets to dominate the response.

To further avoid redundancy, we introduce a diversity loss that encourages distinct activation behaviors. Let $\mathbf{A} \in \mathbb{R}^{B\times K}$ be the activation matrix over the training batch $B$, where each row corresponds to an input instance and each column to a shapelet. To encourage diverse activations across shapelets, we compute the pairwise absolute cosine similarity matrix $\mathbf{C} \in \mathbb{R}^{K\times K}$ as $\mathbf{C}=|\frac{\mathbf{A}^{\top} \mathbf{A}}{||\mathbf{A}||_2^2}|$. The diversity loss is then defined by penalizing the off-diagonal similarity:
\begin{equation}
\mathcal{L}_{\text{div}} = \frac{1}{K(K-1)} \sum_{i \neq j} C_{i,j},
\end{equation}
where $C_{i,j}$ is the cosine similarity between shapelet $\mathbf{S}_i$ and $\mathbf{S}_j$. Minimizing this loss encourages each shapelet to focus on distinct discriminative features, leading to a more compact and expressive representation.
Overall, the full loss function combines the classification loss with the regularization terms:
\begin{equation}
    \mathcal{L}_{\text{total}} = \mathcal{L}_{\text{cls}} + \lambda_1 \cdot \mathcal{L}_{\text{spr}} + \lambda_2 \cdot \mathcal{L}_{\text{div}},
\end{equation}
where $\lambda_1$ and $\lambda_2$ are hyperparameters to control the contributions of sparsity and diversity losses, respectively. In this paper, we set $\lambda_1$ and $\ \lambda_2$ to $0.0001$ by default.

Algorithm~\ref{algo} presents the pseudocode for the end-to-end joint representation learning pipeline. For each mini-batch, the LLM extracts global feature vectors, while the shapelet network computes local activation vectors based on pairwise distances. These activations are projected into local representations and combined with global features to create a joint representation. The model is optimized by minimizing a cross-entropy loss with two regularization terms. 
The overall training is efficient, as only a small subset of parameters is updated. For example, for the GPT-2 base model, all trainable components account for only about $0.5\%$ of the total parameters.

\begin{algorithm}[ht]
    \caption{End-to-end mini-batch training with shapelet-based regularization}\label{algo}
    \begin{algorithmic}[1]
    \renewcommand{\algorithmicrequire}{\textbf{Input:}}
     \renewcommand{\algorithmicensure}{\textbf{Output:}}
     \REQUIRE Training dataset $\mathcal{D} = \{(\mathbf{x}_i,y_i)\}_{i=1}^N$, batch size $B$, input embedding $f_{\alpha}^e$, frozen LLM $f_{\beta}^l$, shapelet network $\psi_\theta$, local linear projection $f_{\phi}^p$, output linear projection $f_{\varphi}^p$, learning rate $lr$, regularization coefficients $\lambda_1$ and $\lambda_2$.
     \ENSURE  Trained learnable modules $f_{\alpha}^e, f_{\beta}^l, \psi_\theta, f_{\phi}^p, f_{\varphi}^p$. \\
     \FOR{number of training epochs}
     \FOR {each mini-batch $\{(\mathbf{x}_i, y_i)\}_{i=1}^B$} 
        \STATE $\mathbf{z}_i^{(g)} \gets f_{\beta}^{l}(f_{\alpha}^{e}(\mathbf{x}_i))$
        \FOR{each shapelet $\mathbf{S}_k$}
                \STATE $d_{i,j}^k \gets ||\mathbf{S}_k - \mathbf{x}_{i,j}^k||_2$
                \STATE $w_{i,j}^k \gets \frac{\exp(-d_{i,j}^k)}{\sum_{j'} \exp(-d_{i,j'}^k)}$
                \STATE $a_i^k \gets \sum_j w_{i,j}^k \cdot (-d_{i,j}^k)$
        \ENDFOR
      \STATE $\mathbf{a}_i \gets [a_i^1, \dots, a_i^K]$
      \STATE $\mathbf{z}_i^{(l)} \gets f_\phi^p(\mathbf{a}_i)$
      \STATE $\mathbf{z}_i \gets \mathbf{z}_i^{(g)} \oplus \mathbf{z}_i^{(l)}$
      \STATE $\mathcal{L}_{\text{cls}} \gets -\frac{1}{B} \sum_{i=1}^{B} \log p(y_i | f_\varphi^p(\mathbf{z}_i))$
      \STATE $\mathcal{L}_{\text{spr}} \leftarrow \frac{1}{B} \sum_{i=1}^{B} \| \mathbf{a}_i \|_1$
      \STATE $\mathbf{A}_{\text{norm}} \leftarrow \text{RowNormalize}(\mathbf{A})$
      \STATE $\mathbf{C} \gets |\mathbf{A}_{\text{norm}}^\top \mathbf{A}_{\text{norm}}|$
      \STATE $\mathcal{L}_{\text{div}} = \frac{1}{K(K-1)} \sum_{i \ne j} C_{i,j}$
      \STATE $\mathcal{L} \leftarrow \mathcal{L}_{cls} + \lambda_1 \cdot \mathcal{L}_{spr}+ \lambda_2 \cdot \mathcal{L}_{div}$
      \ENDFOR \\
      \STATE $\{\alpha, \beta, \theta, \phi, \varphi\} \leftarrow \{\alpha, \beta, \theta, \phi, \varphi\} - lr\cdot \nabla \mathcal{L}$ \\
    \ENDFOR \\
     \RETURN $f_{\alpha}^e, f_{\beta}^l, \psi_\theta, f_{\phi}^p, f_{\varphi}^p$
    \end{algorithmic}
\end{algorithm}

\subsection{Inference Mechanism}
The system supports both standard and few-shot inference paradigms, enabling flexible adaptation to unseen domains with minimal or no labeled data. Both inference paths operate on the joint representation extracted from I/Q signals, which integrates explicit local structure from the shapelet network and global semantics from the pre-trained LLM.

\subsubsection{Standard Inference}
In the standard scenario, the system is evaluated on target domain data without access to any labeled examples from that domain. The system directly processes unseen input samples through the well-trained modules to produce device class predictions. For an unseen input data $\mathbf{x}_j' \in \mathcal{D}'$ from the target domain, the final prediction is:
\begin{equation}
    \hat{y} = f_\varphi^{p}(f_{\beta}^{l}(f_{\alpha}^{e}(\mathbf{x}_j')) \oplus f_\phi^{p}(\psi_{\theta}(\mathbf{x}_j'))).
\end{equation}
This approach enables immediate deployment in new environments without requiring target domain training data by leveraging the pre-trained LLM's generalization capabilities and the learned shapelet patterns to classify devices.

\subsubsection{Few-shot Inference}
Following~\cite{li2025radio}, we formulate cross-domain device authentication with few target domain examples as a few-shot learning problem.
In this case, we enhance the model’s performance through in-context learning, leveraging the LLM’s ability to utilize limited examples to make informed predictions. This process adopts a prototype-based strategy~\cite{snell2017prototypical}, where each class is represented by a prototype vector computed from the support set. 
Given a support set $\mathcal{D}_s' = {(\mathbf{x}_j^s, y_j^s)}_{j=1}^{N_s}$ containing $N_s$ examples per class, the model first extracts joint representations for all samples. For each class, the prototype $\mathbf{c}_k$ is computed by averaging the representations of its corresponding support instances:
\begin{equation}
\mathbf{c}_k = \frac{1}{n_c} \sum_{\mathbf{x}_j^s \in \mathcal{D}_{s,c}'} \mathbf{z}_j, \quad \mathbf{c}_k \in \mathbb{R}^{d_h+d_l},
\end{equation}
where $\mathcal{D}_{s,c}'$ denotes the set of support samples labeled with class $c$, and $n_c$ is the number of such samples. For a query sample $\mathbf{x}^q$, its joint representation $\mathbf{z}_q$ is compared to each class prototype using a similarity metric, and the predicted label corresponds to the most similar prototype:
\begin{equation}
\hat{y} = \arg\max_k \ \text{sim}(\mathbf{z}_q, \mathbf{c}_k),
\end{equation}
where $\text{sim}(\cdot)$ is the negative Euclidean distance in this paper to evaluate the similarity. This few-shot approach enables rapid adaptation to unseen domains with minimal labeled data, effectively utilizing the learned joint representation space.



\section{Experimental Evaluation and Analysis}\label{sec:ex_eva}

\subsection{Experiment Setup}
In all experiments, the learning rate and max epochs are set to $0.0001$ and $200$.  
All experiments are conducted on NVIDIA A100 GPUs with 40GB of memory.

\subsubsection{Datasets}\label{sub:datasets}

This paper employs four public datasets and two self-collected datasets, encompassing Wi-Fi, LoRa, and BLE. Table~\ref{table:dataset_summaries} provides a summary of these datasets. ORACLE~\cite{sankhe2019oracle} uses $16$ USRP X310 transmitters, following the 802.11a standard, with measurements at various distances. Four different distances are selected as unseen domains. The dataset from~\cite{hanna2020open} includes $163$ devices operating on 802.11g. We use data from $58$ devices across five days and refer to this subset as CORES. Three days are used as unseen domains. Collected by the same team as the CORES, the WiSig dataset~\cite{hanna2022wisig} captures signals from $174$ commercial Wi-Fi cards using 802.11a/g on channel $11$ over four days. Given the dataset’s large scale, we only extract data from a single receiver (``node3-19") and treat two of the four days as unseen domains. \cite{elmaghbub2021lora} captures LoRa transmissions from $25$ Pycom devices across five days, which are dubbed as NetSTAR. We use data from two days as target domains.

As shown in Fig.~\ref{fig:lora_device}, our LoRa dataset uses $10$ LoRa transmitters (Pycom LoPy4) and a USRP N210 receiver at four locations, with source domains in line-of-sight (LOS) settings and the target domain in non-line-of-sight (NLOS). The BLE dataset comprises signals from $9$ devices (Nordic nRF52840 Dongle) at different locations, with one LOS and three NLOS as target domains.
These datasets vary across time and location, making them suitable for evaluating domain shift. To standardize model input, all signals are downsampled to a fixed size of $2 \times 256$.

\begin{table}[ht]
    \begin{minipage}{0.66\columnwidth}
        \caption{Summary of Employed Datasets.}
        \label{table:dataset_summaries}
        \centering
        \resizebox{\textwidth}{!}{
        \begin{tabular}{l|ccc}
        \toprule
        \multirow{2}{*}{\textbf{Dataset $\downarrow$}} & \textbf{\# of} & \textbf{\# of} & \textbf{\# of Unseen} \\
        & \textbf{Samples} & \textbf{Devices} & \textbf{Domains} \\
        \midrule
        ORACLE~\cite{sankhe2019oracle}    & 192,000   & 16   & 4 \\
        CORES~\cite{hanna2020open}        & 250,681   & 58   & 3 \\
        WiSig~\cite{hanna2022wisig}       & 270,616   & 130  & 2 \\
        NetSTAR~\cite{elmaghbub2021lora}  & 68,200    & 25   & 2 \\
        LoRa                              & 64,000    & 10   & 1 \\
        BLE                               & 10,800    & 9    & 4 \\
        \bottomrule
        \end{tabular}}
    \end{minipage} 
    \hfill
    \begin{minipage}{0.32\columnwidth}
    \centering
    \begin{figure}[H]
        \includegraphics[width=\linewidth]{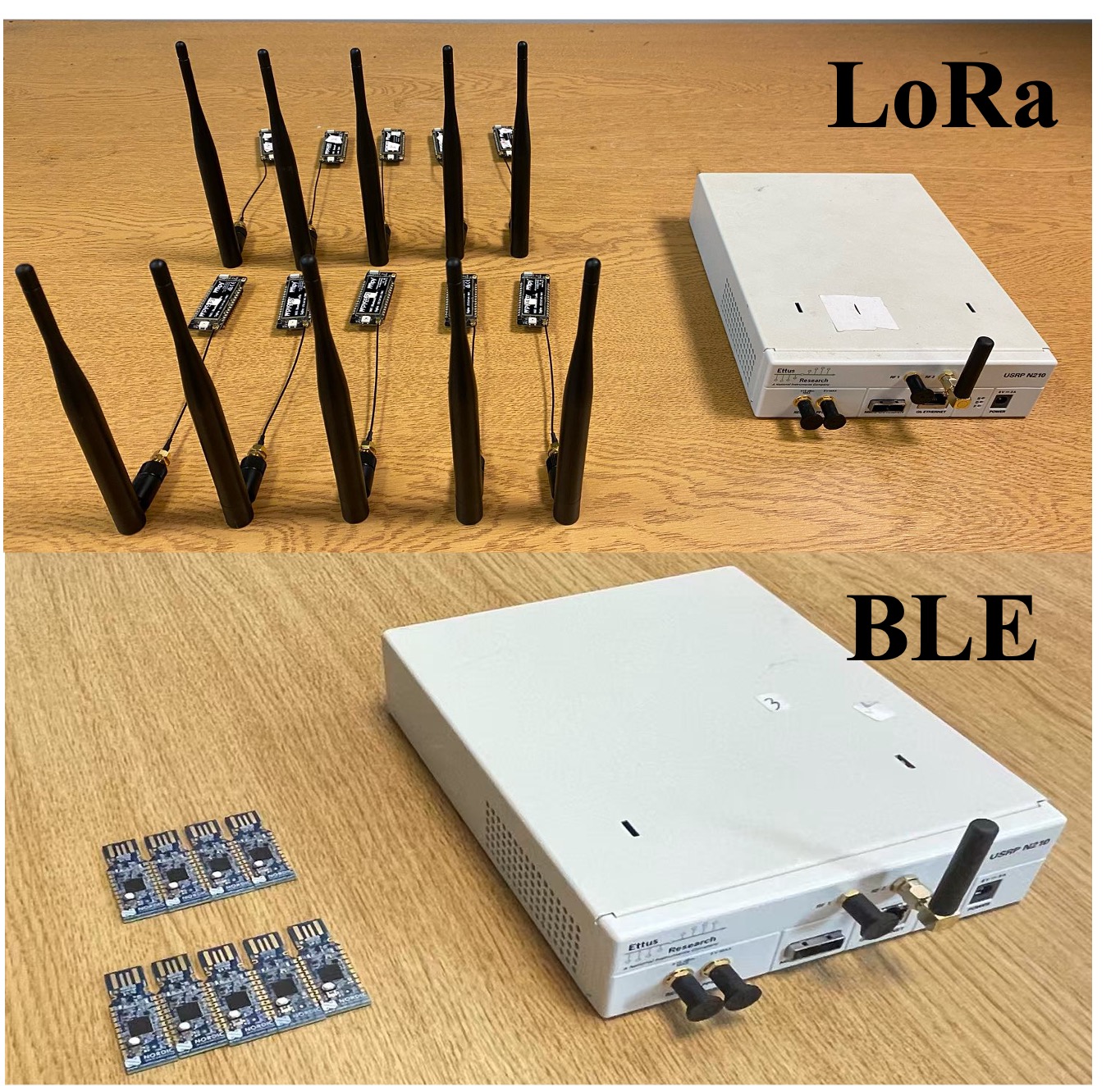}
        \caption{LoRa and BLE dataset devices.}
        \label{fig:lora_device}
    \end{figure}        
    \end{minipage}
\end{table}

\subsubsection{Model Configuration}
The input embedding module $f_\alpha^{e}$ employs a lightweight two-layer CNN with a fixed kernel size of $5$ to transform raw I/Q data into a sequence of length $l_{seq} = 64$, ensuring compatibility with the LLM input format. We use GPT-2 base as the default LLM. The shapelets configuration is $\{(5,8),(5,16),(3,32)\}$. Both projection heads, $f_\phi^{p}$ and $f_\varphi^{p}$, are implemented as single-layer perceptrons. Specifically, $f_\phi^{p}$ projects shapelet features to a $64$-dimensional space, while $f_\varphi^{p}$ maps the joint representation to the number of classes.

\subsubsection{Baseline Methods}
To evaluate the effectiveness of our method under domain shift, we compare it with several baseline methods spanning supervised learning, domain adaptation, SSL, LLM-based modeling, and FSL.
For supervised learning, we choose ResNet-18~\cite{he2016deep} and modify its first layer for I/Q data. For domain adaptation, we include RadioNet~\cite{li2022radionet}, designed specifically for RF fingerprinting under domain shift. In the SSL setting, we adapt LIMU-BERT\cite{xu2021limu}, a BERT-based model originally developed for IMU data, to the RF domain by adapting it for I/Q data. We also include SimCLR~\cite{chen2020simple}, which is implemented using a ResNet-18 backbone for contrastive learning. We also compare with the LLM-based method~\cite{zhou2023one}, which uses patching for input compatibility and a frozen LLM for feature extraction, referred to as PatchLLM.
For FSL, we compare against the customized PTN proposed in~\cite{zhao2024cross}, referred to as RF-PTN. To ensure a fair comparison across all models, we disable any form of target-domain fine-tuning, including the unseen few-shot retraining in RF-PTN.

\begin{table*}[t]
\centering
\caption{Average accuracy of standard inference on source and unseen domains. Best results are highlighted in bold.}
\resizebox{\textwidth}{!}{
\begin{tabular}{c|cc|cc|cc|cc|cc|cc}
\toprule
\textbf{Dataset $\rightarrow$} & \multicolumn{2}{c|}{ORACLE}       & \multicolumn{2}{c|}{WiSig}        & \multicolumn{2}{c|}{CORES}        & \multicolumn{2}{c|}{NetSTAR}      & \multicolumn{2}{c|}{LoRa}         & \multicolumn{2}{c}{BLE}           \\ \hline
\textbf{Method $\downarrow$}  & Source          & Target          & Source          & Target          & Source          & Target          & Source          & Target          & Source          & Target          & Source          & Target          \\ \midrule
ResNet-18~\cite{he2016deep}        & 0.7633          & 0.0743          & 0.9579          & 0.6628          & 0.9948          & 0.7921          & 0.8887          & 0.0997          & 0.5750          & 0.0240          & 0.8212          & 0.7005          \\
RadioNet~\cite{li2022radionet}         & 0.4083          & 0.0611          & 0.8873          & 0.5121          & 0.9537          & 0.6052          & 0.6256          & 0.0296          & 0.5043          & 0.0790          & 0.5950          & 0.1114          \\
LIMU-BERT~\cite{xu2021limu}        & 0.6510          & \textbf{0.2231} & 0.8388          & 0.6423          & 0.7832          & 0.6348          & 0.8504          & 0.0459          & 0.6375          & 0.1295          & 0.6210          & 0.7017          \\
SimCLR~\cite{chen2020simple}           & 0.3054          & 0.1771          & 0.8695          & 0.5700          & 0.9506          & 0.7076          & 0.4554          & 0.0519          & 0.1317          & 0.1050          & \textbf{0.8250} & 0.1481          \\
RF-PTN~\cite{zhao2024cross}              & 0.7218          & 0.0773          & 0.7842          & 0.6517          & \textbf{1.0000} & 0.7714          & \textbf{0.9079} & 0.1939          & \textbf{0.7264} & 0.1083          & 0.6003          & 0.7540          \\
PatchLLM~\cite{zhou2023one}         & \textbf{0.9023} & 0.1189          & 0.9475          & 0.5902          & 0.9963          & 0.7250          & 0.6737          & 0.0245          & 0.6600          & 0.1040          & 0.6625          & 0.3844         \\
Ours             & 0.7129          & 0.1777          & \textbf{0.9749} & \textbf{0.7585} & 0.9998          & \textbf{0.8787} & 0.8507          & \textbf{0.2363} & 0.6795          & \textbf{0.1995} & 0.7444          & \textbf{0.7945} \\ \bottomrule
\end{tabular}}
\label{table:overall}
\end{table*}

\begin{table}[th]
\centering
\caption{$1$-shot and $5$-shot accuracy on unseen domains.}
\resizebox{\columnwidth}{!}{
\begin{tabular}{cc|cccccc}
\toprule
\multicolumn{2}{c|}{\textbf{Dataset $\rightarrow$}}                        & ORACLE & WiSig  & CORES  & NetSTAR & LoRa   & BLE    \\ \midrule
\multicolumn{1}{c|}{\multirow{2}{*}{\textbf{1-shot}}} & RF-PTN  & 0.6579 & 0.8619 & 0.9738 & 0.2820  & 0.8440 & 0.8042 \\
\multicolumn{1}{c|}{}                                 & Ours & 0.7376 & 0.8359 & 0.9769 & 0.2656  & 0.8808 & 0.8207 \\ \midrule
\multicolumn{1}{c|}{\multirow{2}{*}{\textbf{5-shot}}} & RF-PTN  & 0.6983 & 0.9174 & 0.9843 & 0.4315  & 0.8604 & 0.8367 \\
\multicolumn{1}{c|}{}                                 & Ours & 0.8400 & 0.9098 & 0.9954 & 0.4124  & 0.9429 & 0.8511 \\ \bottomrule
\end{tabular}}
\label{table:fewshot}
\end{table}

\subsection{Evaluation on Generalization}
As discussed in Section~\ref{sec:formulation}, a key objective of this work is to improve the generalization of RF fingerprinting models. 
To assess this, we use classification accuracy as the performance metric, which reflects the inverse of the expected error $\mathcal{E}$.

\subsubsection{Cross-domain Evaluation}
We first evaluate the model's performance under cross-domain settings, where the training and testing data are drawn from different domains within the same dataset. This setting reflects realistic deployment scenarios in which domain shifts occur without introducing new devices.
Table~\ref{table:overall} presents the standard inference performance across multiple datasets and baseline methods. Our method consistently achieves the highest accuracy on unseen domains while maintaining strong performance on the source domain. Notably, on the WiSig dataset, our method not only yields the best target-domain accuracy but also improves source-domain performance by approximately $9\%$ compared to the strongest baseline. These results indicate that the proposed method effectively improves generalization across domains while preserving source domain performance. This aligns with the generalization objective defined in Section~\ref{sec:formulation}.

We further assess the model’s adaptability in the few-shot setting, where a small number of labeled samples from the target domain are provided during inference. Unlike traditional fine-tuning approaches, our method leverages the in-context learning capabilities of LLMs to identify devices without any parameter updates. 
Table~\ref{table:fewshot} shows the few-shot inference results across multiple datasets. 
We compare our method with RF-PTN under a conventional few-shot classification setup. For each class, $30$ query instances are used for evaluation.
Each evaluation is repeated $30$ times for stability. We consider two support configurations: $1$-shot and $5$-shot, where each class in the support set is provided with only one or five labeled examples, respectively. These support samples simulate low-resource conditions and are used to construct class prototypes for inference without retraining.
In general, our method demonstrates superior few-shot performance. Especially for ORACLE, our method achieves $73.76\%$ accuracy in the $1$-shot case, a remarkable $56\%$ improvement over the $0$-shot case. This substantial gain highlights the model’s strong in-context learning ability. Additionally, our method outperforms RF-PTN by approximately $8\%$ in the same $1$-shot setting, underscoring its advantage over existing methods.

\begin{figure}[th]
    \centering
    \subfloat[1-shot performance]{%
        \includegraphics[width=0.48\columnwidth]{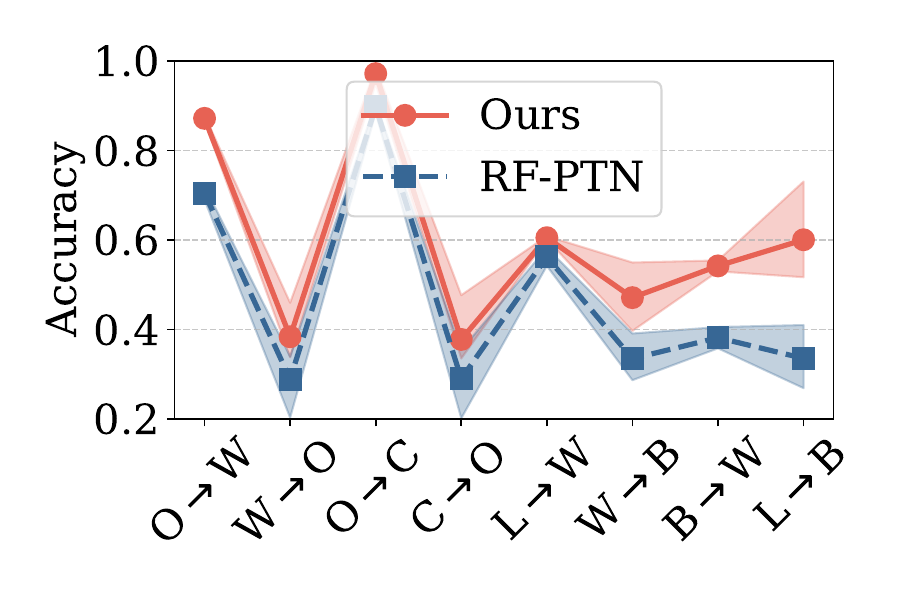}%
        \label{fig:1shot_cross}
    }
    \subfloat[5-shot performance]{%
        \includegraphics[width=0.48\columnwidth]{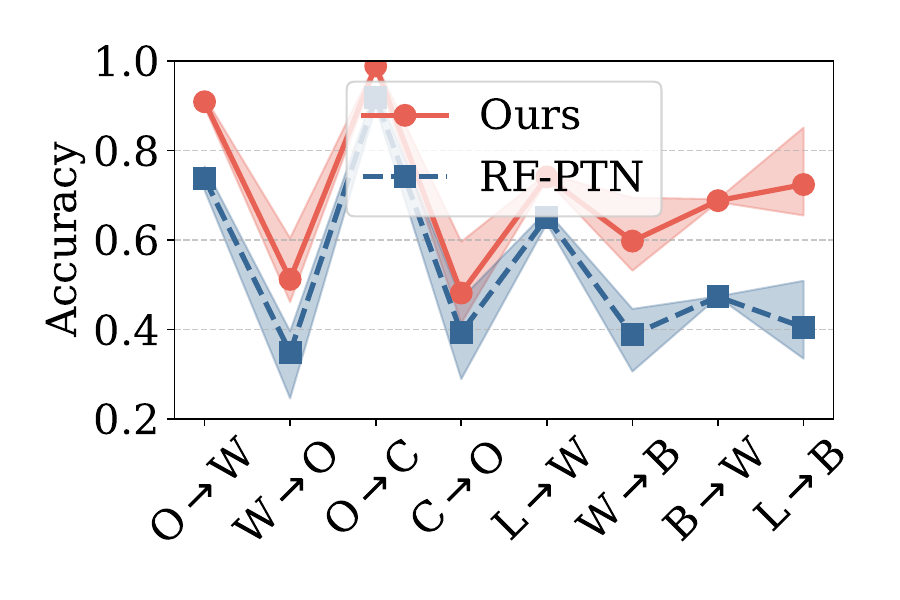}%
        \label{fig:5shot_cross}
    }
    \caption{Few-shot performance in cross-dataset evaluation. The shaded regions represent the accuracy ranges across multiple unseen domains. O, W, C, L, and B refer to ORACLE, WiSig, CORES, LoRa, and BLE, respectively; O$\rightarrow$W indicates training on ORACLE and testing on WiSig.}
    \label{fig:cross_dataset}
\end{figure}

\subsubsection{Cross-dataset Evaluation}
We also assess the model's generalization capability in cross-dataset scenarios, where RF fingerprinting systems encounter new devices not present during training. By leveraging the few-shot inference ability, we efficiently adapt to new devices or even new protocols without retraining. As shown in Fig.~\ref{fig:cross_dataset}, our method demonstrates superior generalization across all cross-dataset scenarios for both $1$-shot and $5$-shot settings. The results consistently outperform RF-PTN across most scenarios, with our lowest-performing domain often surpassing RF-PTN's best. Same-protocol transfers generally yield higher accuracy than cross-protocol transfers, but our method maintains competitive performance even in challenging cross-protocol scenarios. For example, we can achieve about $85\%$ accuracy in the best case LoRa-to-BLE transfer with only five labeled samples.

\begin{figure}[th]
    \centering
    \subfloat[CORES]{%
        \includegraphics[width=0.98\columnwidth]{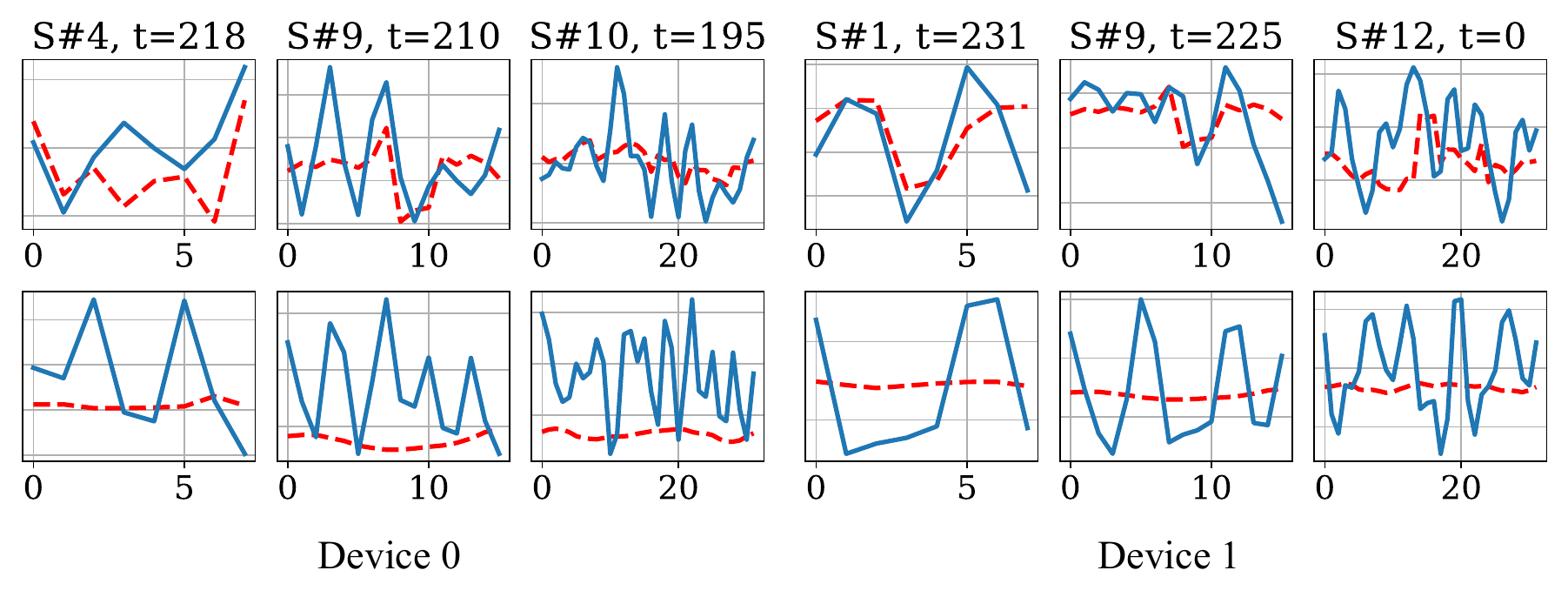}%
        \label{fig:cores_explain}
    }
    \\
    \subfloat[BLE]{%
        \includegraphics[width=0.98\columnwidth]{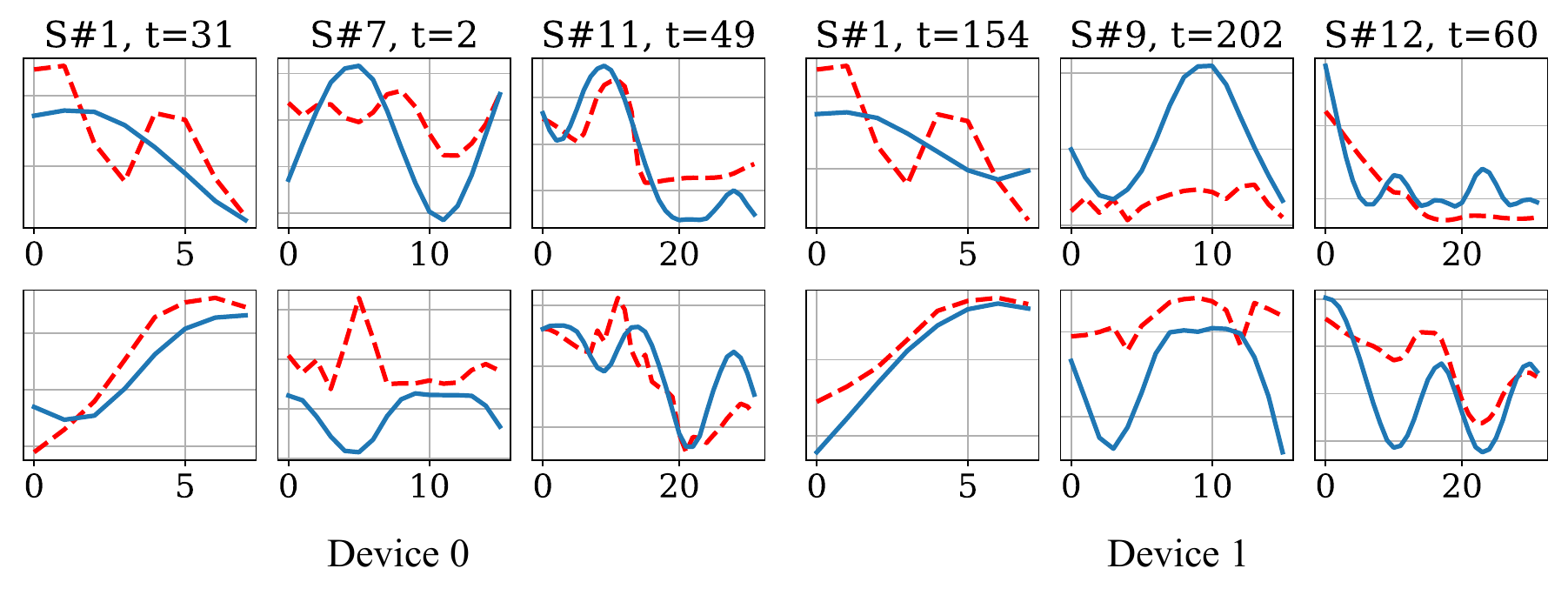}%
        \label{fig:ble_explain}
    }
    \caption{Visualization of learned shapelets and the matched subsequences. S$\#$ denotes the shapelet index, and $t$ indicates the starting time index. The I and Q components are shown in the first and second rows. Blue: real subsequence; red dashed: matched shapelet.}
    \label{fig:explanation}
\end{figure}

\subsection{Evaluation on Interpretability}\label{sec:eval_interpretability}
In addition to generalization, model interpretability is critical in RF fingerprinting to understand which signal components contribute most to device identification. In our framework, interpretability is achieved through a set of variable-length shapelets that offer intrinsic explanations by highlighting discriminative local patterns.
Fig.~\ref{fig:explanation} visualizes selected shapelets and their corresponding matched subsequences. Different devices activate different combinations of shapelets, reflecting device-specific patterns.
On the CORES dataset, most shapelets align with the I component, while the Q component is rarely used. This may be because classification on CORES is relatively easy, and the model can rely solely on the I component. In contrast, for BLE, shapelets match well with both I and Q components, indicating that both are necessary to capture discriminative features for accurate classification. Besides, shorter shapelets tend to fit better, possibly because fingerprint features are localized within short signals.

\begin{figure}[th]
    \centering
    \subfloat[LoRa: 1 domain]{%
        \includegraphics[width=0.32\columnwidth]{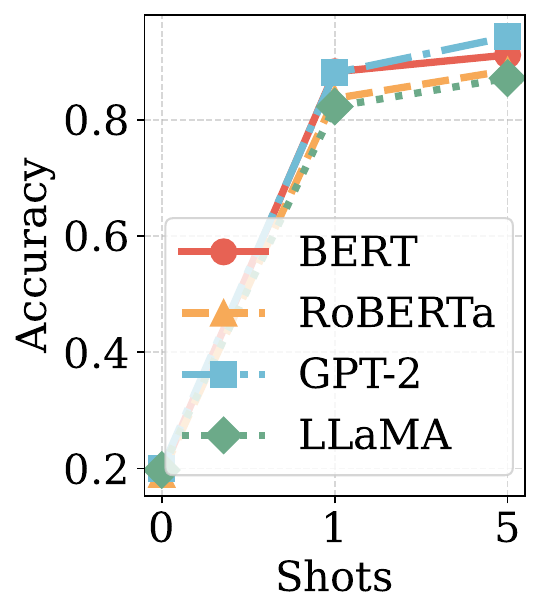}%
        \label{fig:lora_llm}
    }
    \subfloat[WiSig: 2 domains]{%
        \includegraphics[width=0.32\columnwidth]{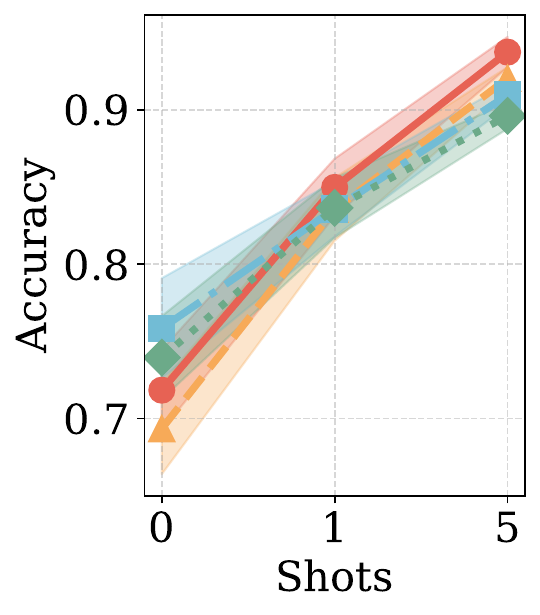}%
        \label{fig:wisig_llm}
    }
    \subfloat[BLE: 4 domains]{%
        \includegraphics[width=0.32\columnwidth]{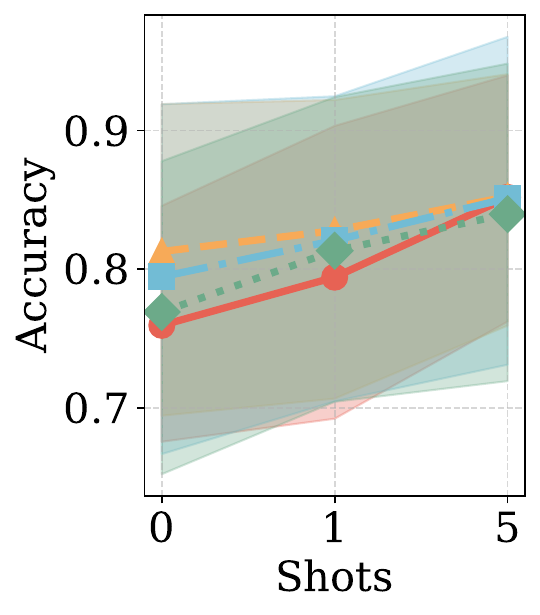}%
        \label{fig:ble_llm}
    }
    \caption{Standard and few-shot performance of using different LLMs. ``0" denotes standard inference. The shaded regions represent the accuracy ranges across multiple unseen domains. LoRa has only one target domain.}
    \label{fig:llms}
\end{figure}


\subsection{Evaluation on Different LLMs}

We further study the impact of different pre-trained LLM backbones in our framework, including GPT-2, BERT, RoBERTa~\cite{liu2019roberta}, and LLaMA~\cite{touvron2023llama}. Fig.~\ref{fig:llms} reports both standard inference (0-shot) and few-shot performance across three protocols. Across all settings, increasing the number of shots consistently improves accuracy, and the performance gaps among different LLMs remain small. In particular, on LoRa and WiSig, all backbones quickly reach comparable accuracy under one-shot supervision.

Interestingly, despite having substantially more parameters, LLaMA does not yield clear advantages over smaller backbones. This suggests that overall performance is not primarily limited by backbone capacity under our lightweight adaptation and fusion design. 
Moreover, the larger hidden size of LLaMA may require more careful feature scaling or fusion calibration with the shapelets module to fully exploit its capacity.
Overall, the consistently similar trends across backbones demonstrate that our method is stable and largely model-agnostic with respect to the choice of pre-trained LLM.


\subsection{Evaluation on Shapelet Module}

\begin{figure}[th]
    \centering
    \subfloat[Source domains]{%
        \includegraphics[width=0.48\columnwidth]{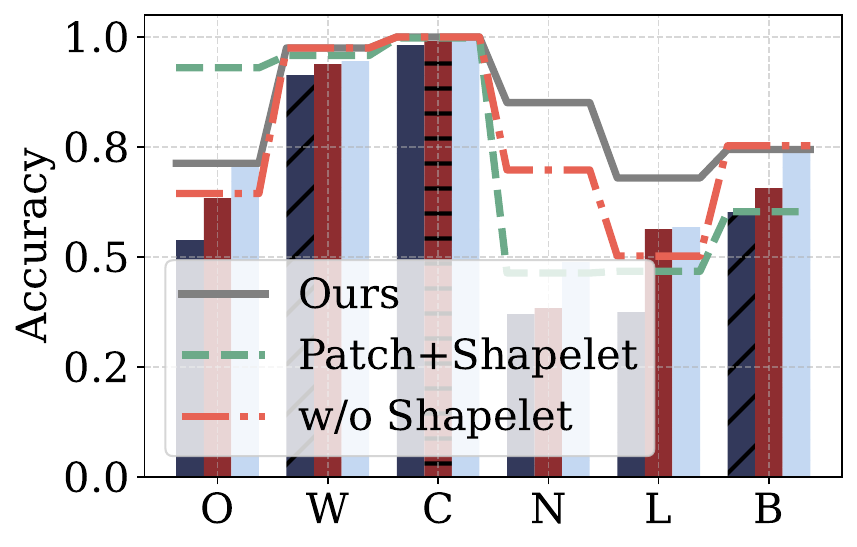}%
        \label{fig:shapelet_source}
    }
    \subfloat[Target domains]{%
        \includegraphics[width=0.48\columnwidth]{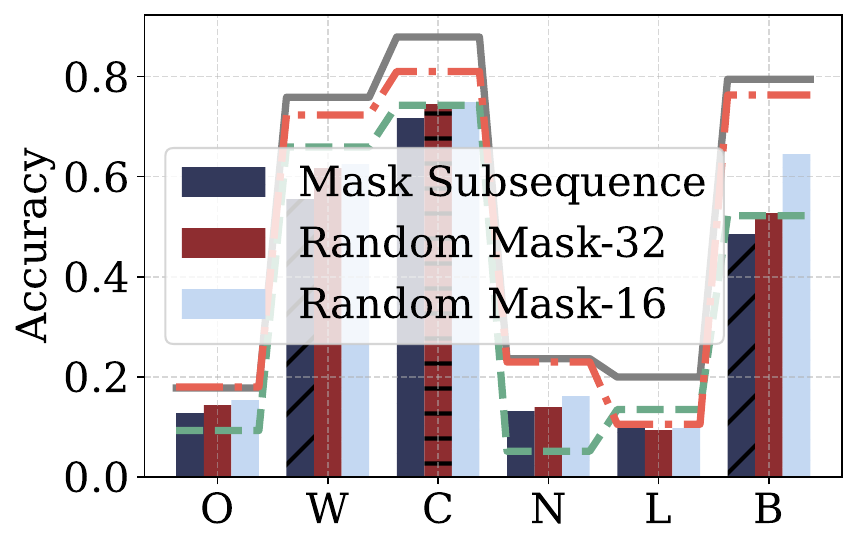}%
        \label{fig:shapelet_target}
    }
    \caption{Shapelets and input embedding module evaluation. O, W, L, and B refer to ORACLE, WiSig, Lora, and BLE, respectively.}
    \label{fig:shapelet_ablation}
\end{figure}

Beyond highlighting local signal patterns, we further examine whether the learned shapelets provide faithful explanations of the model’s decisions. 
As shown in Fig.~\ref{fig:shapelet_ablation}, removing the shapelet module consistently degrades performance on both source and target domains, indicating that shapelets contribute essential discriminative cues for generalization.

To assess causal importance, we mask the subsequence aligned with the highest-activation shapelet ($L\in{8,16,32}$) and compare it with random masking of the same length. 
Across all datasets, masking the shapelet-matched subsequence leads to larger accuracy degradation than random masking. This consistent gap demonstrates that the identified subsequences correspond to decision-critical evidence rather than incidental correlations, confirming the faithfulness of the proposed shapelet-based interpretations.

\begin{table}[th]
\centering
\caption{Comparison of computational cost and parameter efficiency. Both PatchLLM and our method use the same pre-trained BERT-base backbone for fair comparison. Despite the large backbone, our method maintains a comparable computational cost to SOTA methods while minimizing trainable parameters.}
\resizebox{\columnwidth}{!}{
\begin{tabular}{l|c|ccc}
\toprule
\textbf{Model} & \textbf{FLOPs (G)} & \textbf{Total Params (M)} & \textbf{Trainable Params (M)} & \textbf{Trainable Ratio (\%)} \\ \midrule
ResNet-18 & 0.26 & 11.24 & 11.24 & 100.00 \\
PatchLLM & 10.90 & 86.14 & 0.54 & 0.62 \\
\textbf{Ours} & 10.13 & 86.31 & 0.70 & 0.81 \\ \bottomrule
\end{tabular}}
\label{table:cost}
\end{table}

\subsection{Evaluation on Computation Cost}
We evaluate efficiency from two complementary perspectives: (i) computational cost, measured by FLOPs, and (ii) adaptation cost, measured by the number and ratio of trainable parameters. Table~\ref{table:cost} reports results for a fully-trainable CNN baseline (ResNet-18), a frozen-LLM baseline (PatchLLM), and our method.

Although our model incorporates a large pre-trained LLM backbone with 86.31M parameters, only 0.70M parameters ($0.81\%$) are updated during training. This is substantially fewer than the fully-trainable ResNet-18 baseline, which optimizes all 11.24M parameters, demonstrating that our approach enables lightweight adaptation without full-model fine-tuning.
In terms of computation, our method requires 10.13G FLOPs, which is comparable to and slightly lower than PatchLLM (10.90G FLOPs). Since both methods share the same backbone, this indicates that the proposed learnable shapelets module introduces negligible computational overhead while providing additional robustness and interpretability benefits.

Overall, our method improves accuracy and interpretability without sacrificing efficiency. This is achieved by updating only a small fraction of parameters while maintaining similar computational complexity, which reduces optimization cost and memory footprint and makes large pre-trained backbones practical for real-world RF deployment.


\subsection{Ablation Study}

\subsubsection{Input Embedding}
To adapt RF data to LLMs, we employ a CNN-based input embedding module. In time series tasks, patching is a common method to convert time series data to be compatible with Transformers. However, as shown in Fig.~\ref{fig:shapelet_ablation}, replacing our embedding with patching generally leads to performance degradation across source and target domains, indicating the effectiveness of our design for RF data.


\subsubsection{Loss Function}
Table~\ref{table:loss} presents the average performance on source and unseen target domains. Removing the regularization terms typically leads to performance degradation. In particular, $10$ cases exhibit a significant drop, which are highlighted in bold. Among the three components, omitting the sparsity loss $\mathcal{L}_{\text{spr}}$ causes the smallest accuracy drop but still impacts overall performance. These results demonstrate the effectiveness of our loss design in improving performance.

\begin{table}[ht]
\centering
\caption{Impact of loss components on generalization across domains. Bold values indicate significant accuracy drops ($\geq 7\%$).}
\resizebox{\columnwidth}{!}{
\begin{tabular}{cc|cccccc}
\toprule
\multicolumn{2}{c|}{\textbf{Dataset}}                      & ORACLE          & WiSig  & CORES  & NetSTAR         & LoRa            & BLE             \\ \midrule
\multicolumn{1}{c|}{\multirow{2}{*}{w/o $\mathcal{L}_{\text{div}}$}}  & $\mathcal{D}$ & \textbf{0.6378} & 0.9747 & 0.9998 & 0.8383          & \textbf{0.5033} & 0.7583          \\
\multicolumn{1}{c|}{}                             & $\mathcal{D}'$ & 0.1342          & 0.7354 & 0.8511 & 0.2272          & \textbf{0.0997} & \textbf{0.7241} \\ \midrule
\multicolumn{1}{c|}{\multirow{2}{*}{w/o $\mathcal{L}_{\text{spr}}$}}  & $\mathcal{D}$ & \textbf{0.6291} & 0.9756 & 0.9999 & {0.8012} & \textbf{0.4883} & 0.7694          \\
\multicolumn{1}{c|}{}                             & $\mathcal{D}'$ & 0.1720          & 0.7456 & 0.8625 & 0.2478          & 0.2000          & 0.7725          \\ \midrule
\multicolumn{1}{c|}{\multirow{2}{*}{only $\mathcal{L}_{\text{cls}}$}} & $\mathcal{D}$ & \textbf{0.6034} & 0.9690 & 0.9999 & \textbf{0.7156} & \textbf{0.4042} & 0.7653          \\
\multicolumn{1}{c|}{}                             & $\mathcal{D}'$ & 0.1552          & 0.7410 & 0.8599 & 0.2337          & \textbf{0.0787} & 0.7614          \\ \bottomrule
\end{tabular}}
\label{table:loss}
\end{table}

\section{Limitation and Future Work}\label{sec:future}
\subsection{Deployment Efficiency}
Although this paper leverages pre-trained LLMs to achieve superior performance without cumbersome retraining procedures, it still relies on substantial computational and storage resources when deployed in a server-side setting. In particular, the frozen LLM backbone incurs non-trivial memory footprints and inference latency, which may limit scalability in resource-constrained environments.
In this paper, our framework assumes centralized processing with sufficient compute and storage budgets. Extending this approach to edge or on-device deployment remains an open challenge, as it would require significantly more efficient designs and hardware-aware optimization techniques. Promising future directions include model compression and distillation for frozen LLMs, sparse shapelet selection to reduce inference overhead, and collaborative edge–server inference schemes that balance accuracy and efficiency.
Such advances would not only improve deployability but also pave the way for scaling to larger and more powerful foundation models in practical RF systems.

\subsection{Adaptive Shapelet Design}
Our current shapelet design employs multiple fixed-length windows to capture RF fingerprint patterns at different temporal scales. While this multi-scale strategy improves representational coverage, it inevitably introduces a trade-off between flexibility, efficiency, and performance. In particular, using a small set of predefined window sizes may lead to redundant features across overlapping scales, while still failing to optimally capture device-specific fingerprint structures that deviate from these fixed lengths.
A promising direction for future work is to develop adaptive windowing strategies that automatically adjust both shapelet lengths and the number of shapelets in a data-driven manner. Such designs could mitigate feature redundancy, improve representational efficiency, and more faithfully match shapelet granularity to the intrinsic temporal structure of RF fingerprints.

\subsection{Physical-Level Interpretability}
While the proposed shapelet-enhanced framework improves interpretability by highlighting salient temporal segments that influence classification, the explanations remain primarily at the representation level. The identified shapelets indicate where and when discriminative patterns occur, but are not yet explicitly connected to the underlying physical-layer imperfections that generate RF fingerprints (e.g., hardware nonlinearity or clock offsets).
Establishing such links would enable more physically grounded and causal interpretations beyond segment-level saliency. We consider this a promising direction for bridging learned representations with domain knowledge in wireless systems.

\section{Conclusion}\label{sec:conclusion}

In this paper, we propose a novel approach to mitigate domain shift and enhance interpretability in DNN-based RF fingerprinting. Our method adapts pre-trained LLMs to improve cross-domain generalization and integrates variable-length shapelets to provide intrinsic explanations of model predictions. The training is guided by a combination of classification, sparsity, and diversity losses to encourage discriminative and interpretable representations. Moreover, we employ prototype-based inference to leverage the few-shot capability of LLMs, enabling rapid adaptation with minimal labeled data. Extensive experiments across diverse domains and protocols demonstrate the effectiveness and robustness of our approach.


\bibliographystyle{ACM-Reference-Format}
\bibliography{LM}


\end{document}